\documentclass{article}

\usepackage{arxiv}

\usepackage[utf8]{inputenc} 
\usepackage[T1]{fontenc}    
\usepackage{hyperref}       
\usepackage{url}            
\usepackage{booktabs}       
\usepackage{amsfonts}       
\usepackage{nicefrac}       
\usepackage{microtype}      
\usepackage{graphicx}
\usepackage{natbib}
\usepackage{doi}

\usepackage{amsmath,amssymb,amsfonts}
\usepackage{algorithmic}
\usepackage{graphicx}
\usepackage{textcomp}
\usepackage{supertabular,booktabs}
\usepackage{lscape}
\usepackage{amsmath}
\usepackage{setspace}
\usepackage{subcaption}

\title{An Affective Video Database using Multimedia Content Analysis rated on Indian samples}

\author{ {\hspace{1mm}Sudhakar Mishra} \\
	Department of Information Technology\\
	Indian Institute of Information Technology Allahabad\\
	Prayagraj, IN 211012 \\
	\texttt{rs163@iiita.ac.in} \\
	\And
	{\hspace{1mm}Narayanan Srinivasan} \\
	Department of Cognitive Science\\
	Indian Institute of Technology Kanpur\\
	Kanpur, IN 208016 \\
	\texttt{nsrini@iitk.ac.in} \\
 	\And
	{\hspace{1mm}Uma Shanker Tiwary} \\
	Department of Information Technology\\
	Indian Institute of Information Technology Allahabad\\
	Prayagraj, IN 211012 \\
	\texttt{ust@iiita.ac.in} \\
}

\renewcommand{\shorttitle}{\textit{arXiv} Template}

\hypersetup{
pdftitle={A template for the arxiv style},
pdfsubject={q-bio.NC, q-bio.QM},
pdfauthor={David S.~Hippocampus, Elias D.~Striatum},
pdfkeywords={First keyword, Second keyword, More},
}

\begin{document}
\maketitle

\begin{abstract}
	Availability of naturalistic affective stimuli is needed for creating the affective technological solution as well as making progress in affective science. Although a lot of progress in the collection of affective multimedia stimuli has been made in western countries, the technology and findings based on such monocultural datasets may not be scalable to other cultures. Moreover, the available dataset on affective multimedia content has some experimenter bias in the initial manual selection of affective multimedia content. Hence, in this work, we mainly tried to address two problems. The first problem relates to the experimenter's subjective bias, and the second relates to the non-availability of affective multimedia dataset validated on Indian population. We tried to address both problems by reducing the experimenter's bias as much as possible. We adopted the data science and multimedia content analysis techniques to perform our initial collection and a further selection of stimuli. Our method resulted in a dataset with a wide variety in content, stimuli from Western and Indian cinema, and symmetric presence of stimuli along valence, arousal and dominance dimensions. We conclude that using our method, more cross-cultural affective stimuli datasets can be created, which is essential to make progress in affective technology and science. 
\end{abstract}


\section{Introduction}
\label{sec:introduction}

The availability of a standardized affective video database is not only needed for technological advances in affective computing but required for scientific advances in the field of affective science also. In the technological domain, despite the rapid improvement in visual object detection and recognition in computer vision, the affective technology is yet in the primitive state, waiting for sincere research engagement and availability of affective content to analyze \cite{van2011affective}. Likewise, the scientific progress of affective science is also staggered at the point where excellent previous works with static stimuli left \cite{saarimaki2021naturalistic}. Technology and science are struggling to consider the stimulus's context so that better scientific theories and affective technological solutions can be developed. Moreover, the cultural dominance over the scientific theory of affects and emotions obstructs the path of generalization. For example, although western research on affective science is very progressive and inspiring, struggling with the issue of applicability of the results in Western, educated, industrialized, rich, and democratic societies (WEIRD) \cite{henrich2010weirdest}. Hence, a database with cultural diversity in mind is needed to propose a probable solution to the above-stated problem. 

Although significant progress has been made in the availability of affective content, the datasets currently available (described below) have many limitations. For instance, the initial collection of candidate stimuli in these datasets is biased by the experimenter's selections \cite{holman2015evidence} limiting the variety in the content. Instead, if this initial collection process is automatized and multimedia social platforms are crawled objectively for affective content, the variety in the affective content would be more promising. Another limitation is that some currently available datasets are not downloadable or easily available to the research community due to third-party license or copyright issues. Hence, it is needed to create a dataset which could be easily available and downloadable to the research community.

Another major limitation in datasets with affective multimedia content is that most datasets are dominated by Western content. Although due to the availability of affective multimedia Western content significant technological and scientific progress has been made, the progress may not necessarily translate to other cultures. The so-called Global south is mostly untouched in terms of the development of affective multimedia content. The Indian subcontinent shares a larger share of the world population. However, the availability of data for research is an issue in India \cite{pandey2010health}. Hence, developing the affective technology for a diverse country like India is a challenge which becomes almost impossible if enough data is not developed to analyze the affective dynamics of the Indian population. The widely acceptable datasets currently available for research are described below. 

\subsection*{Affective Multimedia Databases}
Creating an affective stimuli dataset is very much needed to develop the affective computing or recommendation systems model and to perform emotion experiments with naturalistic stimuli that can resemble real-life contexts in lab settings. Although to date few datasets of affective video stimuli have been created, they have mainly three limitations. First, the experimenter selects the stimuli, which may limit the variety in the stimuli set. Second, the datasets also has limitation in the number of emotions assumed or assigned for these stimuli. Finally, all the affective datasets recorded to date predominantly have stimuli from western cinema and lack the inclusion of stimuli from Indian cinema. 

The first film stimuli dataset was created by \cite{philippot1993inducing} and \cite{gross1995emotion} with a small set of stimuli to elicit specific emotional categories in the laboratory. Since then, many efforts have been made to increase the size of affective stimuli datasets further. For instance, Douglas-Cowie et al. create the HUMAINE database \cite{douglas2007humaine}. This dataset had 50 naturalistic and induced data clips with duration ranging from 5 seconds to 3 minutes. These stimuli were labeled with a wide range of emotion labels at the global level.

A FilmStim dataset created by \cite{schaefer2010assessing} consisted of 70 film excerpts with the aim to use them for eliciting emotional states in an experimental psychology experiment. Seventy film excerpts were distributed equally among six emotional states (i.e. anger, sadness, fear, disgust, amusement, tenderness) and one non-emotional state. Three sixty-four participants rated film clips, and based on 24 classification criteria ranking scores were computed \cite{schaefer2010assessing}. 

Another film stimuli dataset was created by \cite{koelstra2011deap} called the DEAP dataset. This dataset has 120 one-minute-long music videos excerpts. At least 14 volunteers rated these excerpts in online mode. The rating scales included valence, arousal, dominance, and emotion labels with intensity. Youtube links for these music videos were provided, but all are not accessible due to either copyright law or removed from YouTube.

MAHNOB-HCI is another dataset created in the same year by \cite{soleymani2011multimodal}. This database has 20 short emotional excerpts from movies and videos. Participants rated the excerpts on the valence and arousal scale. 

A dataset (EMDB) with 52 non-auditory film clips were created by \cite{carvalho2012emotional}. These clips were 40 seconds long and were extracted from commercial films. The clips were selected in such a way that they covered the entire valence-arousal affective space. One hundred thirteen participants rated these clips along valence, arousal and dominance dimensions. Although non-auditory clips facilitate future experiments for audio manipulation, the absence of audio modifies viewers' perception of emotional feelings. In addition, the dataset cannot be used for multimodal processing.

A dataset with violent scenes was created by Demarty et al. \cite{demarty2012benchmarking}. Due to its specific genre, this dataset is primarily used for 'violent scenes detection'. The segments of violent scenes were embedded in 25 movies that elicited highly arousing and negative emotions. Although the movies are annotated for violent scenes, due to copyright issues, movies together with annotations are not easily available. 

A dataset with the short animated GIF is created by researchers in MIT \cite{jou2014predicting}. Due to the popularity of these short videos, especially on YouTube, they are good for eliciting a set of emotions. However, covering a larger range of emotional categories with these short videos and animated GIFs may be difficult.

Recently, a dataset Chieti Affective Action Videos (CAAV) with ninety actions is created \cite{di2020chieti}. The CAAV dataset contains action videos filmed for 15 seconds on actors performing a wide variety of affective and non-affective actions. Each action is filmed in four versions; male, female, first-person perspective and third-person perspective. Four hundred forty-four participants rated these action films on valence and arousal dimensions. The data has a wide variety of affective actions situated in a context. However, the videos are less naturalistic and embed less situational complexity; hence, they might not resemble real-life situations. 

Other than the western database, a Chinese database, Chinese Natural Emotional Audio-Visual Database (CHEAVD), with naturalistic emotional multimedia stimuli, was created \cite{li2017cheavd}. A total of 140 min emotional segments from films, TV plays, and talk shows were rated by 238 native speakers, aged from children to the elderly. This dataset claimed to elicit 26 non-prototypical emotional states, including six basic emotions. 

Recently, a news-based video database is reported to investigate the dynamics of emotion, and memory \cite{samide2020database}. One hundred twenty-six news clips were used to elicit negative, positive and neutral feelings about real-life news events. The dataset contains ratings of valence by participants at the local and global levels. Participants also rated the overall intensity of the news clip. Other than affective ratings, participants also performed a memory test by reporting the video details and duration at the end of each video. The dataset is limited to positive, negative and neutral feelings. 

Another Chinese dataset published recently contains 242 short videos, including 112 positive videos, 47 neutral videos, and 83 negative videos \cite{li2022visual}. The duration of the videos was 3 seconds which seems to be very less to situational contextualize an emotion. Moreover, the dataset contained videos that elicit positive, negative and neutral feelings. 

The datasets mentioned above are mostly dominated by western stimuli, which limits them to be used by researchers investigating the western population only. Although datasets from other cultures came out \cite{li2017cheavd, chen2021selection, li2022visual}, no video stimuli dataset is available for the Indian subcontinent, which shares almost one-fifth of the world population. 

The main contributions of this paper are as follows.
\begin{itemize}
    \item Automatizing the collection of candidate multimedia items to cover a wide variety in content without any subjective preference and biases.
    \item The availability of large set of video excerpts which is downloadable for use by research community.
    \item First video dataset which contains both Western and Indian stimuli validated by Indian participants with proficiency in expressions of all forms of English language.
\end{itemize}



\begin{table*}[!h]
    \centering
    \begin{tabular}{c|c|c|p{7cm}}
    Name & \#clips & Duration & Emotional Labels \\ \hline
    HUMAINE & 50 & 5s to 3min.& Reporting wide range of global states including emotion-related states, context labels, key events, emotion words, etc. Assessment of intensity, valence, arousal, dominance, and predictability at frame-by-frame level. \\
    FilmStim & 70 & 1min. to 7min.& 24 classification criteria: Use of the differential emotions scale to derive valence and arousal affect. \\
    DEAP & 120 & 1min. & Online self-assessment on valence, arousal and dominance. EEG, EMG, \& ECG recording with 40 film stimuli. \\
    MAHNOB-HCI & 20 & 35s to 117s & Ratings of valence, arousal, dominance and emotional keyword. Recording of facial videos, EEG, peripheral physiological signals, gaze and audio. \\
    EMDB & 52 & 40s & Ratings of valence, arousal and dominance dimensions at global level. \\
    VIOLENT SCENES DATASET & 25 & movies & Annotation of movie segment from two different kinds of violent movies. \\ 
    LIRIS-ACCEDE & 9800 & 8s to 12s & Valence and arousal dimensions ranking \\ \hline
    \end{tabular}
    \caption{Downloadable video databases}
    \label{tab:database}
\end{table*}

\section{Method}
\subsection{Stimuli Collection}
\begin{figure*}
    \centering
    \includegraphics{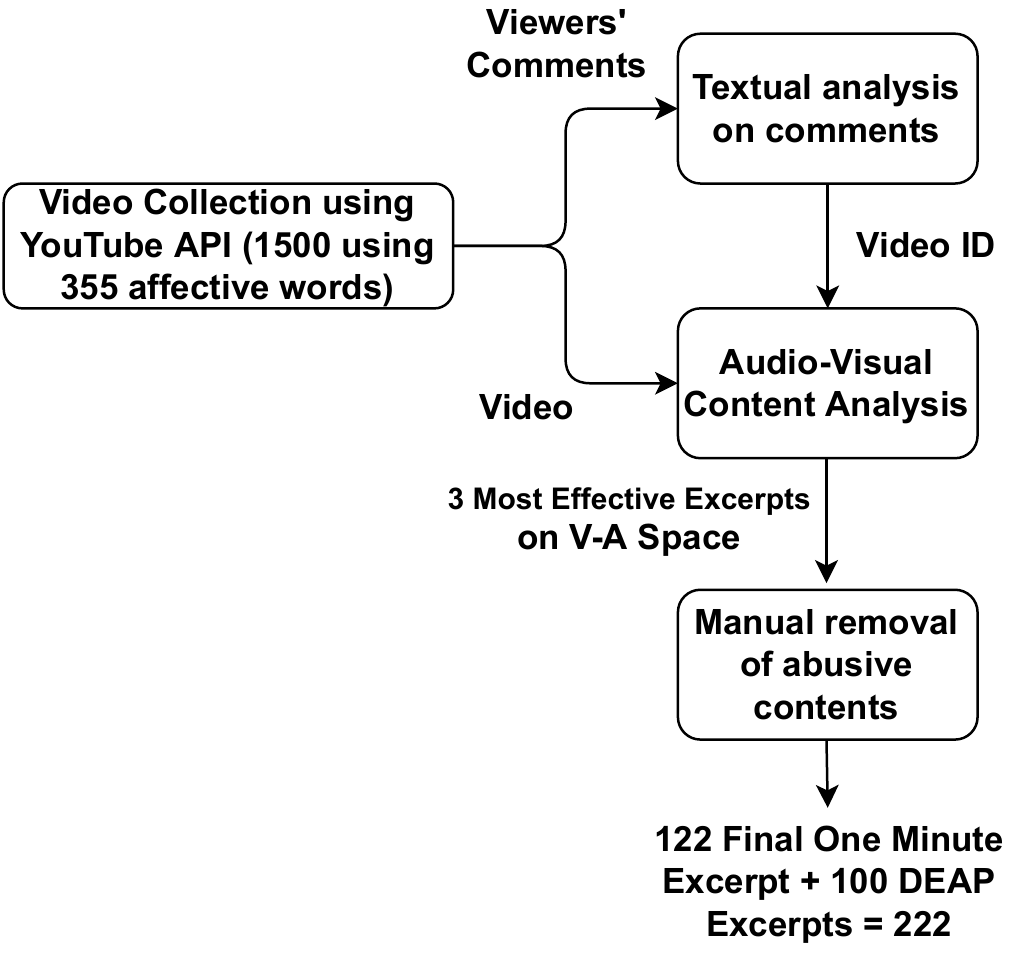}
    \caption{Stimuli Collection Procedure}
    \label{fig:stimCollect}
\end{figure*}

The stimuli collection procedure is depicted in figure-\ref{fig:stimCollect}
\subsubsection*{Affective Keywords}
The list of Affective keywords are taken from WordNet Affect which is an extension of WordNet based on Princeton's English WordNet 3.0. The list of affective words were organized in a tree \cite{affectList}. We created a text file of affective words from the affect tree (available in the GitHub repository \cite{affectList}). The list had 355 affective words. These 355 affective keywords were entered in the YouTube search engine API.

\subsubsection*{Using YouTube search engine API for affective tagging based emotional videos retrieval}
YouTube is a social media platform hosting largest number of user-generated videos on the web. YouTube facilitates searching the content using text words that match the video descriptions, tags, comments and so on. 

We used python's multiprocessing to process YouTube Data API request in parallel \cite{kready2020youtube} and to retrieve the videos. The parallelized YouTube data collection process deals with the wait time for I/O operations which bottlenecks the data collection efforts. The I/O operations at several intermediary steps, including JSON conversions, feature engineering, and data storage, can increase overall wait time if serial processing is being performed. 

An API key through the Google Developers Console has to be created, to use YouTube API. With an API key, a user can make up to 10,000 requests per day. We used the \textit{Request library} in python for making an API request. Request to API can be submitted by creating a URL containing the API URL, API key, Content ID, and data part. Upon successful submission of a request, a JSON response will be created containing the information about the results (e.g. video id) against the query. The video id was further used to extract 100 comments per video. By using the above procedure, we collected 1500 raw multimedia videos along with their comments with YouTube search engine API for affective words taken from Affect word-net \cite{strapparava2004wordnet}.

 \subsection{YouTube Comments Analysis}
One hundred fifty comments per video were fetched. Each comment was given a sentiment score (using the AFINN sentiment analysis \cite{nielsen2011new} and AFINN v0.1 python library) with +1 (positive) and -1 (negative) based on the polarity of the comment. Comments with positive and negative scores were pooled separately to calculate summed up positive and negative polarity of each video. If the positive score is greater than the negative score, then the video was assigned high positive polarity and vice versa (if the negative score was higher). This resulted in 175 videos with positive or negative polarity. These videos were downloaded from YouTube using YouTube API. Additionally, 120 videos of one-minute excerpts from DEAP data were included. In total, 295 videos were analyzed using multimedia content analysis. 

\subsection{Multimedia Features}

Before applying multimedia content analysis, the videos downloaded from YouTube were divided into 60 seconds segments with 55 seconds of overlap between consecutive segments. Multimedia content-based analysis on these videos was performed to calculate audio-visual features of the scene, and these audio-visual features were further used to create a set of fixed duration video clips. The audio-visual features (table-\ref{tab:MultiMediaFeatures}) have been reported in cinematography to induce affective experiences and are presumably linked to different affective dimensions like arousal and valence \cite{audioAnalysis-2, yazdani2013multimedia, affectiveContent-1}. Hence, these features were used to calculate the valence and arousal values of the one-minute excerpts. The features are described below.

\subsubsection*{MFCC}
Mel-Frequency Cepstral Coefficient is a way to represent the signal in terms of coefficients of frequencies converted to mel scale. The core idea of representing the signal in mel spectrum (frequency domain representation) is that humans do not differentiate between lower and higher frequencies. Human ears can perceive the difference between lower frequencies better than the higher ones. That means better resolution in lower frequencies than higher frequencies is the characteristic which should be modelled. However, the traditional frequency domain analysis does not account for this characteristic. Hence, the frequency spectrum is needed to be converted to the mel spectrum using the following steps. First, the signal goes to a pre-emphasis filter to balance the amplitudes in low-frequencies and high-frequencies. Then, fast fourier transform on slices of signals using overlapping windows are created. The obtained frequencies in the frequency spectrogram are used to calculate the filter banks. Discrete cosine transform is applied to the filter banks in order to obtain MFCCs. 

A first order pre-emphasis filter to balance the frequency spectrum is $$y(t) = x(t) - \alpha x(t-1)$$. Here, the filter coefficient $(\alpha = 0.95)$ or 0.97.

Framing is performed using Hamming window as follows
$$w[n] = 0.54-0.46cos(\frac{2\pi n}{N-1})$$

where, $0\leq n \leq N-1$, N is the window length.

N-point FFT is calculated on each frame and then the power spectrum from the FFT frequency components is computed as follows $$P = \frac{|FFT(x_i)|^2}{N}$$
where, $x_i$ is the $i^{th}$ frame of the signal x.

Then, the signal is converted to mel scale from the frequency scale using the following formula $$m = 2595\log_{10} (1+\frac{f}{700})$$

Mel filter banks are calculated by applying triangular kernels on a mel scale. Typically, 40 filter banks are calculated. Triangular kernels have a response of 1 at the center frequency and decrease linearly towards 0. 

The first filter bank will start at first point, will peak at the second point then return to zero at the third point. The second filter bank will start at second point, peak at third point, and will return to zero at fourth point. It can be modelled using following formula.

\begin{equation*}
H_m(k) = \begin{cases} 
0 & \quad k<f(m-1) \\
\frac{(k-f(m-1))}{f(m)-f(m-1)} & \quad f(m-1)\leq k < f(m) \\
\frac{f(m+1)-k}{f(m+1)-f(m)} & \quad f(m) \leq k \leq f(m+1) \\
0 & \quad k > f(m+1)
\end{cases}
\end{equation*}

Apply DCT to decorrelate the filter bank coefficients and get a compressed representation of filter banks in terms of mel frequency cepstarl coefficients.


\subsubsection*{Energy}
Energy of a signal is calculated as 
$$E = \Sigma_{t=-\infty}^{t=\infty}{|x(t)|^2}$$
\subsubsection*{Short term Entropy of Energy}
The short term entropy of energy is calculated for subframes and informs about the abrupt changes in the energy level of a signal. To calculate short term entropy of energy each short-term frame is divided into subframes of fixed duration. The procedure is as follows
$$e_j = \frac{E_{subFrame_j}}{E_{shortFrame_i}}$$
where
$$E_{shortFrame_i} = \Sigma_{k=1}^K{E_{subFrame_k}}$$

Finally the entropy is calculated as $$H(i) = -\Sigma_{j=1}K{e_j.log2(e_j)}$$

\subsubsection*{Spectral Entropy}
Spectral entropy is the calculation of short term entropy of energy in the frequency domain \cite{abreha2014environmental}. The spectrum of the short-term frame is divided into L sub-bands (bins). The normalized energy $n_f$ of each short-term frame is calculated as 

$$n_f = \frac{E_f}{\Sigma_{f=0}^{L-1}{E_f}}, \;\; f\in{0, 1, ...., L-1}$$

The entropy of normalized spectral energy $n_f$ is computed using:
$$H = -\Sigma_{f=0}^{L-1}{n_f.log_2(n_f)}$$

\subsubsection*{Spectral Centroid}
Spectral centroid is calculated as
$$centroid = \frac{\Sigma_{k=b_1}^{b_2}{f_k s_k}}{\Sigma_{k=b_1}^{b_2}{s_k}}$$

where, $f_k$ is the $k^{th}$ bin frequency. $s_k$ is the spectral value at bin k. $b_1$ and $b_2$ are the lower and upper limits of the spectral range over which spectral spread has to be calculated.

\subsubsection*{Spectral Spread}
It is the standard deviation calculated in the frequency domain representation of the signal. The formulation of spectral spread is

$$spread = \sqrt{\frac{\Sigma_{k=b_1}^{b_2} {(f_k-\mu_1)^2S_k}}{\Sigma_{k=b_1}^{b_2}{S_k}}}$$

where, $f_k$ is the $k^{th}$ bin frequency; $S_k$ is the spectral value at bin k. $b_1$ and $b_2$ are the lower and upper limits of the spectral range over which spectral spread has to be calculated. $\mu_1$ is the spectral centroid calculated as described above.
\subsubsection*{Spectral Flux}
Spectral flux quantifies the change in the power spectrum of a signal. It is calculated by taking the squared difference between the power spectrum of consecutive frames.
$$flux(t) = (\Sigma_{k=b_1}^{b_2}{|S_k(t)-S_k(t-1)|^p})^{1/p}$$
where, $S_k$ is the spectral value at bin k. $b_1$ and $b_2$ defines lower and upper cutoff of the band over which flux has to be calculated. $p$ is the norm type.

\subsubsection*{Spectral Roll-off}
The spectral roll-off is the frequency corresponding to the spectrogram bin such that at least roll\_percent of the spectral energy of the considered frame is contained in the current bin and all the bins below the current bin. The calculation of spectral roll-off can be formalized as \cite{abreha2014environmental}
$$SR = f(N)\; where \; f(N) = (\frac{f_s}{K})N$$
where N being the largest bin that fulfills the following condition
$$\Sigma_{k=0}^{N}{|X(k)|^2} \leq TH.\Sigma_{k=0}^{K-1}{|X(k)|^2}$$

\subsubsection*{Chroma vector}
Calculation of chroma representation is useful in extracting the harmonic information of music \cite{bartsch2005audio}. 

Converting the frequency of the calculated spectrogram in the logarithmic scale where the log-frequency is labelled with the nearby pitch frequency. The formalization of the process to derive the pitch based log-frequency spectrogram is as follows:
$$Y_{LF}(m, p) = \underset{k\in P(p)}{\Sigma}{|\chi(m, k)|^2}$$

where P is the set of all MIDI numbered pitches such that 
$$P(p) = \{ k\in [0:K] : F_{pitch}(p-0.5) \leq F_{coef}(k)<F_{pitch}(p+0.5) \}$$

where $F_{coef}$ is the closest pitch with center frequency assigned to the spectral coefficients $\chi(m, k)$ of spectrogram calculated using STFT. $F_{pitch}(p)$ is the center frequency of a pitch $p\in [0:127]$. The center frequency $F_{pitch}(p)$ of a pitch is calculated as 
$$F_{pitch}(p) = 2^{(p-69)/12}.440$$

So, the frequency axis of the STFT based spectrogram is partitioned logarithmically and labelled linearly according to MIDI pitches.

The pitch then can be separated into two components, which are referred to as tone height and chroma. An octave, which is the basic or trivial music interval, defines the interval of pitches. For instance, $p=12$ and $p=24$ are one octave apart and pitches $p=36$ and $p=40$ are two octaves apart. These set of 12 pitches constitute the chroma values and are element of a set $\{C, C^{\#}, D, ...., B\}$ with $c=0$ refers to chroma C, $c=1$ to chroma $C^{\#}$, and so on. The set of all pitches that share the same chroma constitutes the pitch class. For example, pitch class related to $c=1(C^{\#})$ consists of the set $\{1, 13, 25, 37, 49, 61, 73, 85, 97, 109, 121\}$ (making the musical notes $\{C^{\#}0, C^{\#}1, C^{\#}2, C^{\#}3, ....C^{\#}11\}$.)

The chroma feature is the aggregation of all spectral information related to a given pitch class (c) into a single coefficient. Hence,
$$C(m,c) = \underset{\{p\in [0:127]|p \;mod\; 12=c\}}{\Sigma} {Y_{LF}(m,p)} \;\; for \; c\in [0:11]$$

Whereas, the chroma vector (\textbf{C}) $\in [0:11]$. 

\subsubsection*{Chroma deviation}
Chroma deviation differentiates music from environment \cite{wang2006affective}. It is the standard deviation of the 12 chroma coefficients. The chroma difference is defined on chroma vector (\textbf{C}) is $$\Sigma_{i=1}^{11}{|C_{i+1}-C{i}|}$$.

Where, $C_i$ is the $i^{th}$ entry of \textbf{C}.

\subsubsection*{Motion Vector}
We calculated the motion vector using optical flow technique proposed by Lucas-Kanade. Optical flow technique captures the velocity of moving object. Optical flow captures the motion of objects and surfaces using relative motion of the camera. Motion vector for one of the stimulus is shown in figure-\ref{fig:motion}

Mathematically, to know how much a pixel has moved from one location to another in two consecutive time frames is termed as pixel correspondence problem. To solve the problem, two assumptions are made. First assumption is that the motion is small and second assumption is that the appearance doesn't change abruptly between two consecutive time frames. Hence, first assumption allows us to calculate the new position of the pixel in the vicinity of its older position. The second assumption constraints the abrupt change in brightness (brightness constancy constraint) which can be best described as 

$$I(x,y,t) = I(x+u, y+v, t+1)$$

where, $I(x,y,t)$ is the intensity. Taylor series expansion of above equation is 
$$I(x+u, y+v) \approx I(x,y) + \frac{\partial I}{\partial x}u + \frac{\partial I}{\partial x}v$$

Replacing the taylor series expansion in the equation of constancy constraint give us $I_t + I_x u + I_y v = 0$ implies $I_t = -(I_x + I_y)$. That means the spatial motion induced by the camera movement explains changes in the intensity of the pixel. The calculated values of u and v in the x and y direction respectively constitutes the motion vector. 

The calculation of motion vector is not done for each individual pixel due to insufficient information. Hence, it is assumed that collection of colacated pixels will have the same u and v. Simultaneously change in the intensity of these set of pixels is modelled using the least-square estimation which gives us the motion vector and its direction. 

\subsubsection*{Rhythm Components}
The rhythm component is calculated as 
$$G_2(k) = \frac{\underset{k}{max}(c(k))}{\underset{k}{max}(\tilde c(k))} \tilde c(k)\%$$
where $\tilde c(k) = K(l_1, \beta_1) * c(k)$. $K(l_1, \beta_1)$ is the kaiser window (shown in figure-\ref{fig:kaiser}).

$$c(k) = 100 e^{(\frac{1-(n(k)-p(k))}{\delta})}\%$$
where, n(k) and p(k) are the two closest right and left shot boundaries from frame k \cite{hanjalic2005affective}. The curve of calculated rhythm for one stimulus is shown in the figure-\ref{fig:rhythm}

\subsubsection*{Color Energy}
Psychological studies show that brightness is correlated with the emotional valence \cite{wilms2018color}. To calculate the color energy cue the relationship between saturation, and brightness is needed. Equation to calculate the color energy is as follows 
$$\Sigma_i \Sigma_j p(c_i) x p(c_j) x d(c_i, c_j) x \Sigma_k^M{E(h_k)s_k v_k}$$

The histograms of hue, lightness, and saturation are calculated. Together termed as HLS histogram. $c_i$ and $c_j$ are histogram bins indexed with i, j to iterate over all the histogram bins. $d(c_i, c_j)$ is the distance between histogram bins ($L_2$ norm) in HLS space. $p(.)$ is the histogram probability. $s, v, E(h)$ are the saturation, lightness and hue energy respectively of each pixel \cite{wang2006affective}. 

\subsubsection*{Shot rate}
\begin{figure*}

\vspace{-5ex}

\begin{subfigure}{0.49\linewidth}
    \includegraphics[width=\textwidth]{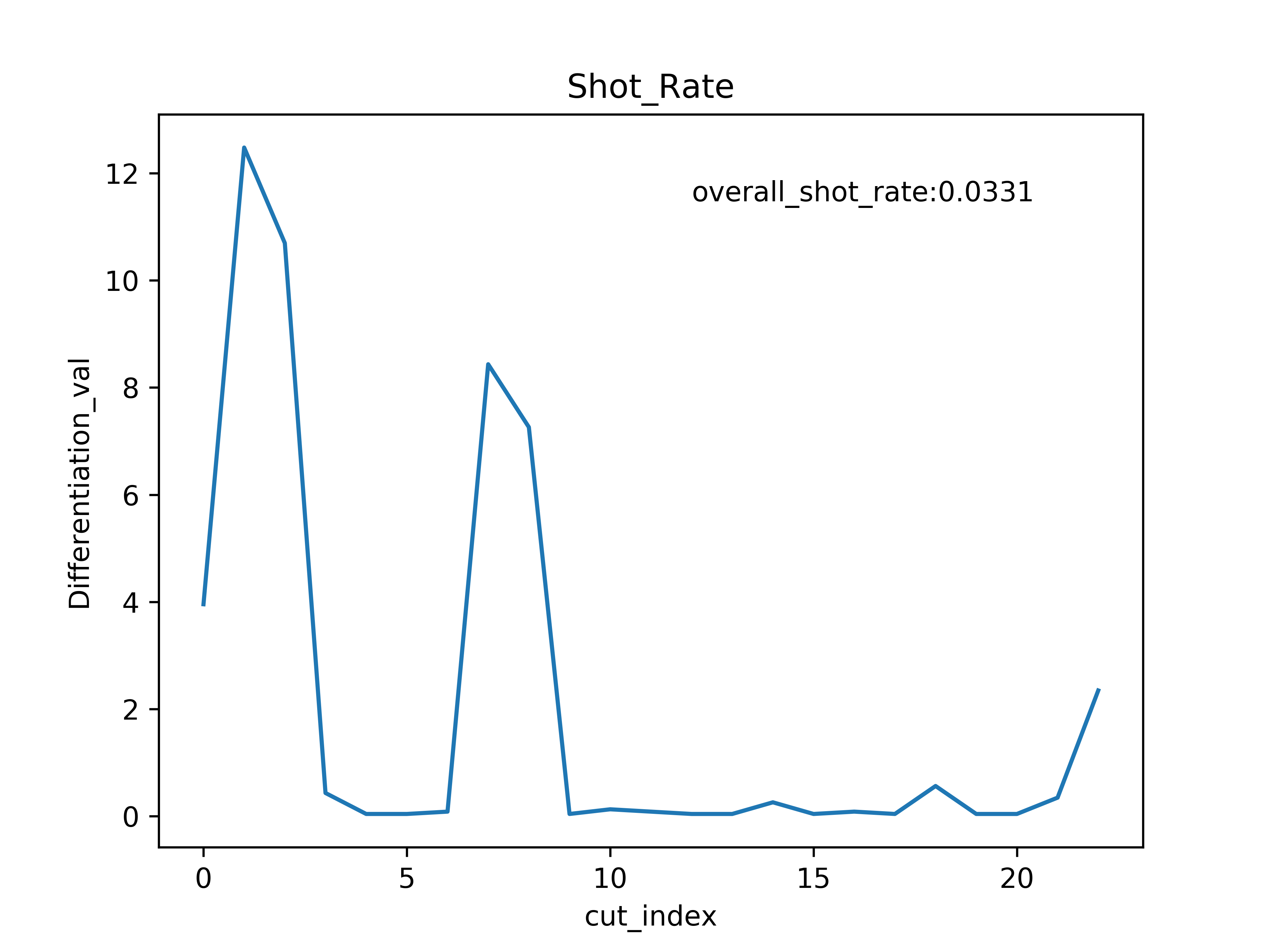}
    \vspace{-1.5\abovedisplayskip}
    \caption{}
    \label{fig:shotRate}
\end{subfigure}
\begin{subfigure}{0.49\linewidth}
    \includegraphics[width=\textwidth]{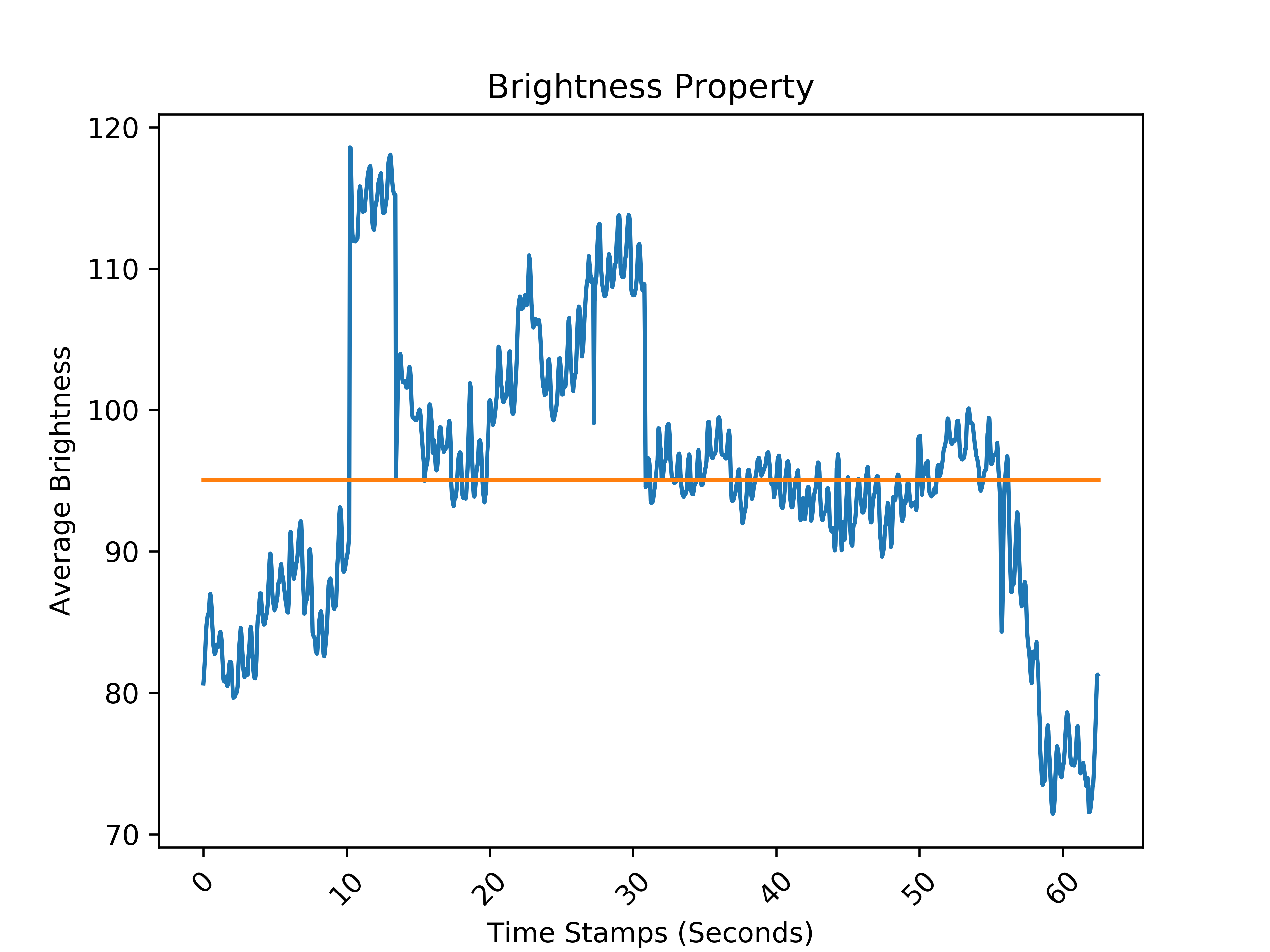}
    \vspace{-1.5\abovedisplayskip}
    \caption{}
    \label{fig:brightness}
\end{subfigure}

\begin{subfigure}{0.49\linewidth}
    \includegraphics[width=\textwidth]{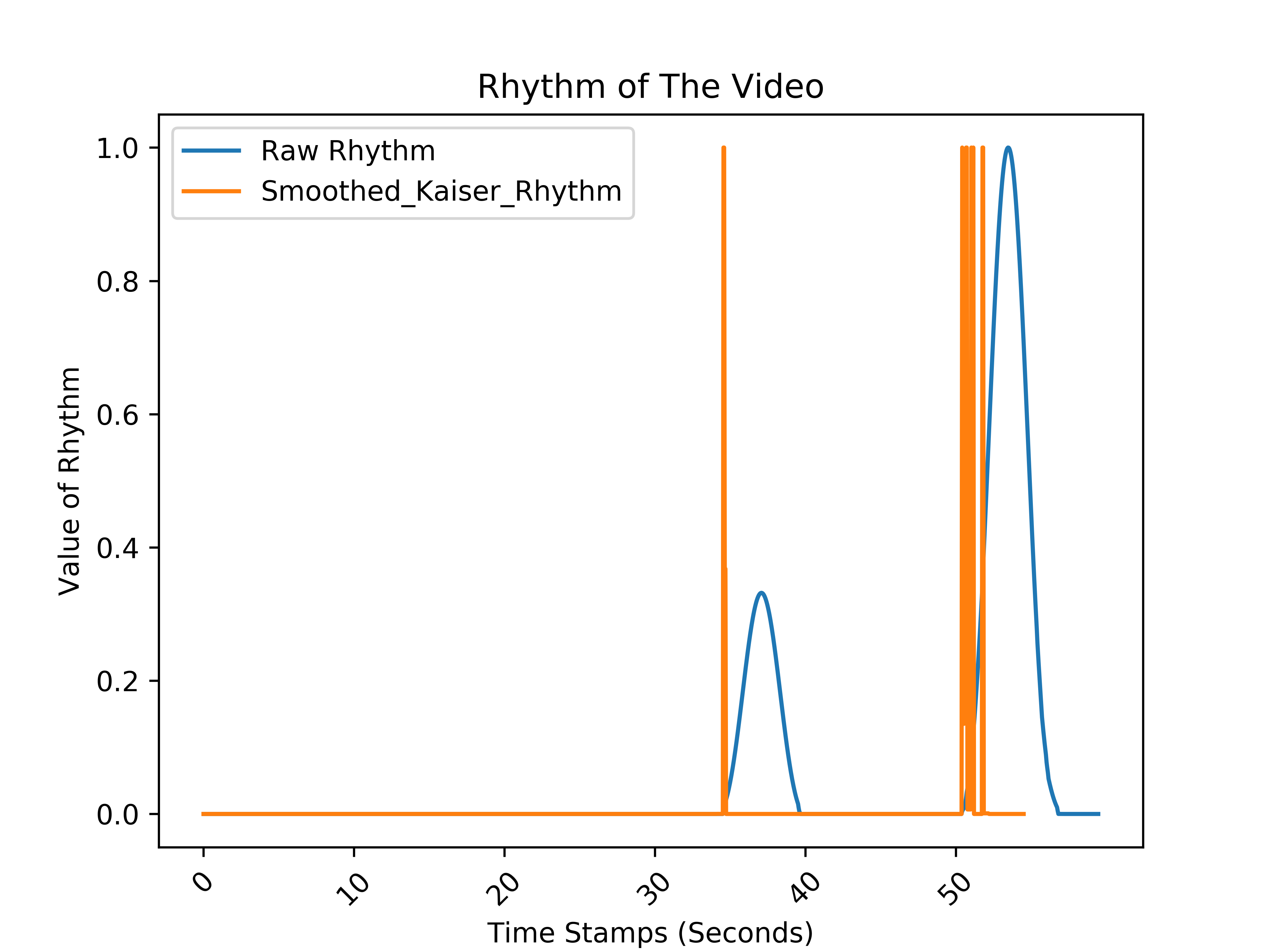}
    \vspace{-1.5\abovedisplayskip}
    \caption{}
    \label{fig:rhythm}
\end{subfigure}
\begin{subfigure}{0.49\linewidth}
    \includegraphics[width=\textwidth]{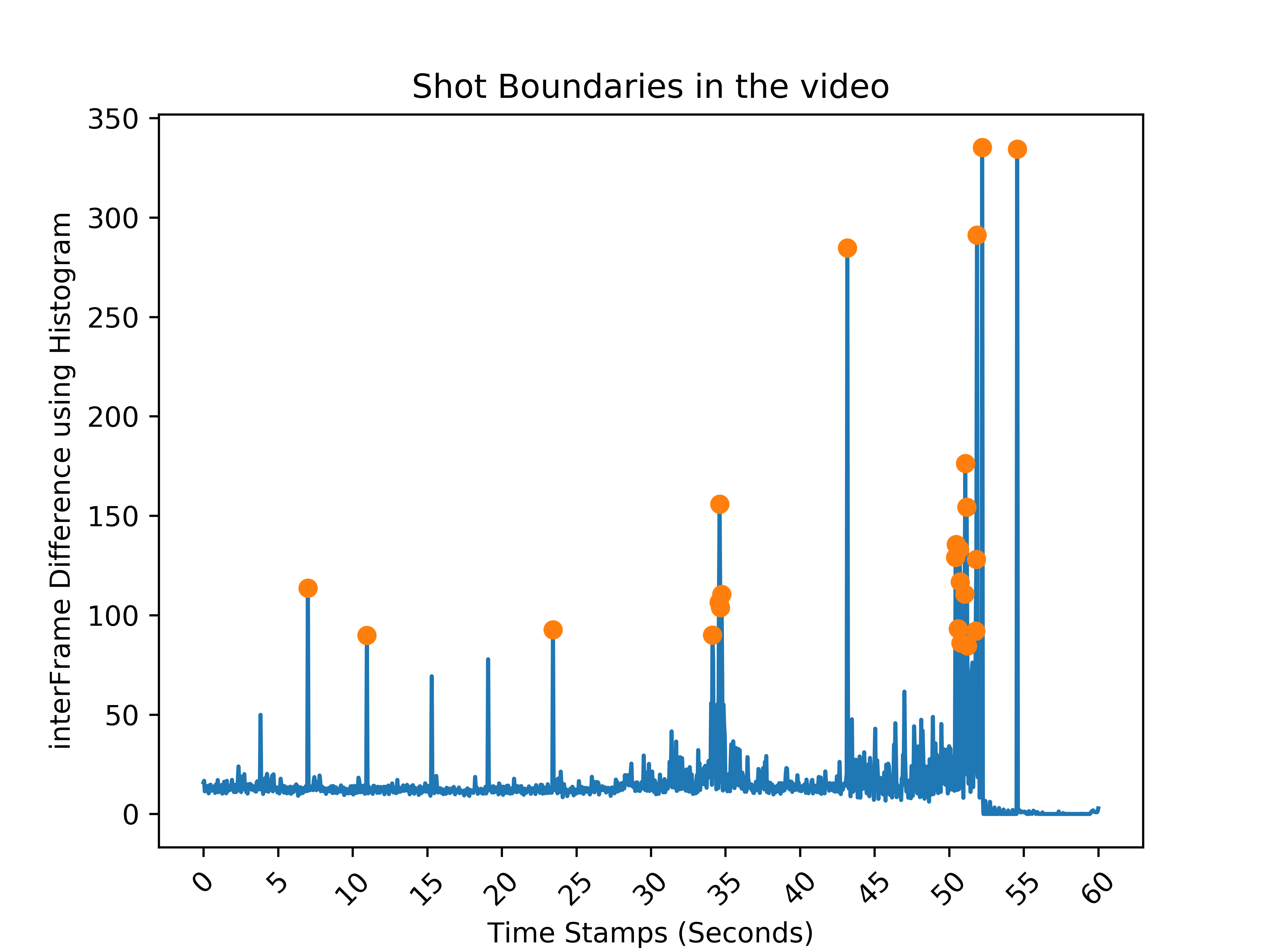}
    \vspace{-1.5\abovedisplayskip}
    \caption{}
    \label{fig:shotBound}
\end{subfigure}

\begin{subfigure}{0.49\linewidth}
    \includegraphics[width=\textwidth]{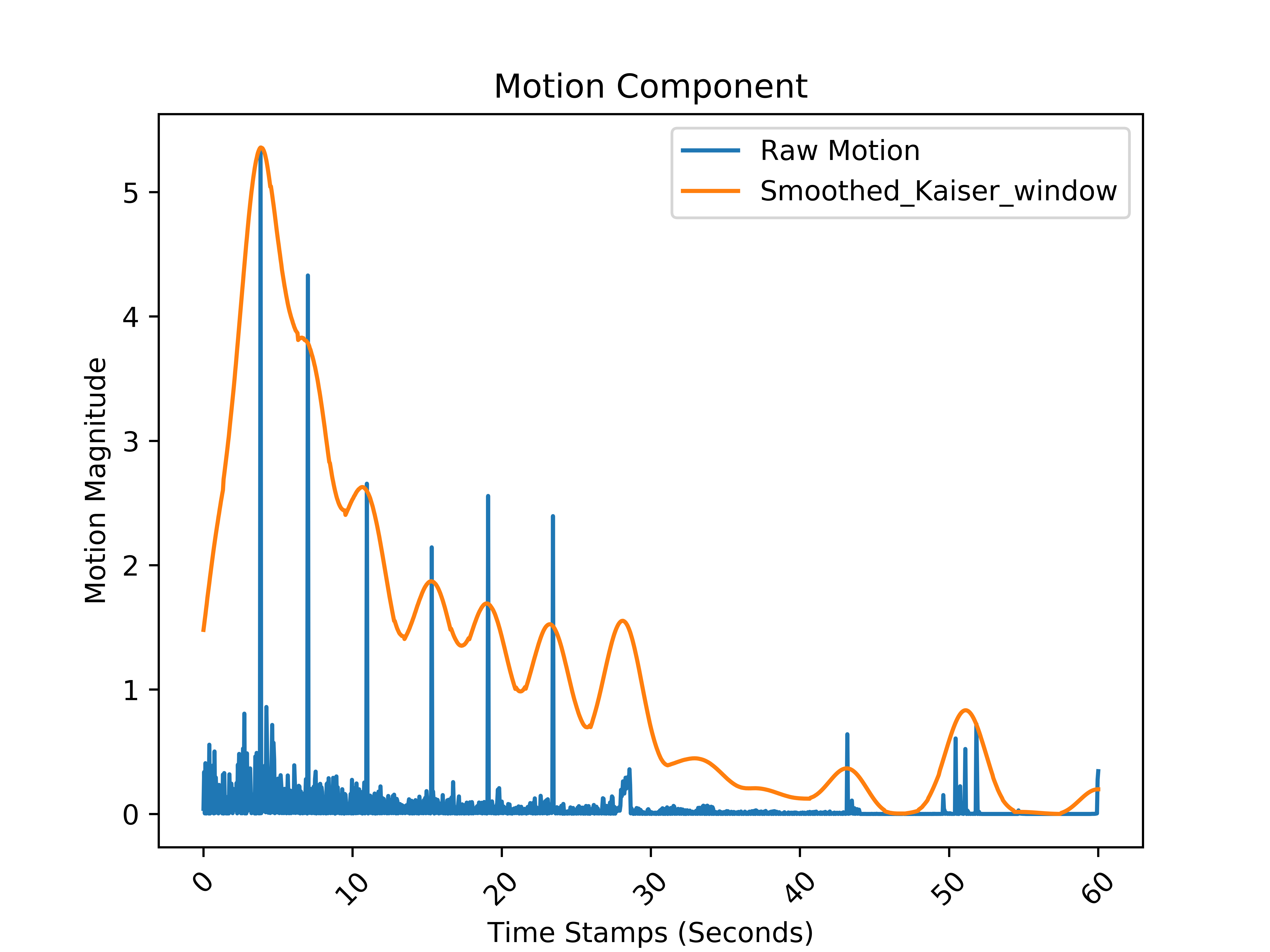}
    \vspace{-1.5\abovedisplayskip}
    \caption{}
    \label{fig:motion}
\end{subfigure}
\begin{subfigure}{0.49\linewidth}
    \includegraphics[width=\textwidth]{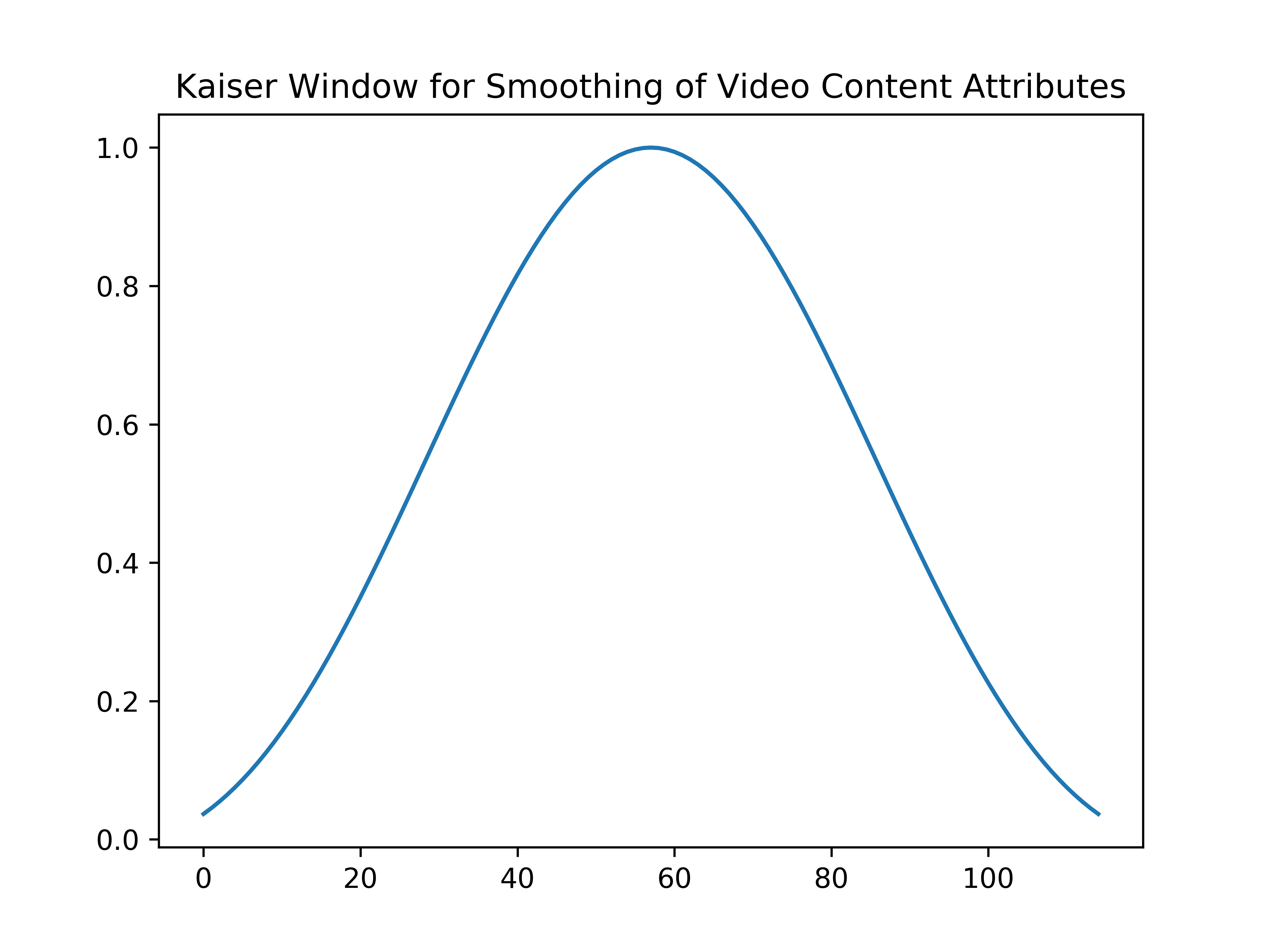}
    \vspace{-1.5\abovedisplayskip}
    \caption{}
    \label{fig:kaiser}
\end{subfigure}
\caption{(a) Shor rate, (b) Brightness, (c) Rhythm (d) Shot Boundaries, (e) Motion curve (f) Kaiser smoothing window}
\label{fig:featureCurves}
\end{figure*}

From the cinematographic perspective, by manipulating the perceived passage of time, the director can heighten arousal. One way to manipulate the pace is by editing effects like cuts which define the shot length. A shot conveys an event, and by rapid shot changes, the density of events can be intensified to heighten the arousal in the viewer \cite{wang2006affective}. For instance, covering the main action from different angles with rapid shot changes convey the dynamic and breathtaking excitement \cite{wang2006affective}. The shot boundaries are needed to be calculated to calculate the shot rate (see figure-\ref{fig:shotBound} for trace of shot boundary). For each pair-wise frame, the shot boundary profile (M) can be computed by calculating the sum of distances (L2-norm) between the L histograms. A frame is declared as the shot boundary if it fulfils the follows criteria
$$C1(i):|\frac{\partial M}{\partial j}|_{j=i} > Th_{b1}$$
$$C2(i):|\frac{\partial^2 M}{\partial j^2}|_{j=i} > Th_{b2}$$
$C3(i): for \; i-15 < j < i+15, C1(j) \;and \; C2(j)$ should both be true only when $j = i$. Both the threshold in above equation ($Th_{b1} \; and Th_{b2}$) are decided experimentally. In a sense, being the first-order equation C1 will detect the frame-to-frame changes only and cannot disambiguate the second-order rapid shot changes. On the contrary, C2 can F true shot boundary from the presence of fast or large moving objects. C3 conditions the maximum perceivable length of shots. Using shot boundaries, shot lengths are calculated, and the rate of change in shot length is the shot rate. The curve of short rate for a stimulus is shown in the figure-\ref{fig:shotRate}.

\begin{table*}[!h]
    \centering
    \begin{tabular}{p{2.5cm}|p{9.6cm}|p{0.5cm}|p{1cm}}
    \toprule
\textbf{Features} & \textbf{Description} & \textbf{V/A} & \textbf{Ref} \\ \hline
    \multicolumn{4}{c}{\textbf{Audio Features}} \\ \hline 
    MFCC-26 & Mel frequency cepstral coefficients from a cepstral representation where the frequency bands are not linear but distributed according to the mel-scale. & V,A & \cite{audioAnalysis-7,audioAnalysis-2} \\ \hline
    Energy-2 & The sum of squares of the signal values, normalized by the respective frame length. & A & \\ \hline
    Spectral entropy-2 & Entropy of the normalized spectral energies for a set of sub-frames & A & \cite{audioAnalysis-2} \\ \hline
    ZCR-2 &The rate of sign-changes of the signal during the duration of a particular frame&A & \cite{audioAnalysis-5,audioAnalysis-6,audioAnalysis-7} \\ \hline
    Short-term entropy of energy-2 &The entropy of sub-frames normalized energies. It can be interpreted as a measure of abrupt changes&  & \cite{audioAnalysis-2} \\ \hline
    Spectral centroid-2 & Spectral centroid is the "center of mass" of the spectrum and represents the brightness of a sound&V & \cite{audioAnalysis-2} \\ \hline
    Spectral spread-2 &The second central moment of the spectrum&V & \cite{audioAnalysis-2} \\ \hline
    Spectral flux-2 &The squared difference between the normalized magnitudes of the spectra of the two successive frames&A & \cite{audioAnalysis-5,audioAnalysis-7,audioAnalysis-2} \\ \hline
    Spectral roll-off-2 &The frequency below which 90\% of the magnitude distribution of the spectrum is concentrated& A & \cite{audioAnalysis-5,audioAnalysis-7,audioAnalysis-2} \\ \hline
    Chroma vector-24 &A 12-element representation of the spectral energy where the bins represent the 12 equal tempered pitch classes of western type music(semitone spacing)&A & \cite{audioAnalysis-2} \\ \hline
    Chroma deviation-2 &The standard deviation of the 12 chroma coefficients & A & \cite{audioAnalysis-2} \\ \hline
     
    \multicolumn{4}{c}{\textbf{Video Features}} \\ \hline
    Motion Vector-7 &Motion vector using Lukas-Kanade algorithm & A & \cite{MotionDetection} \\ \hline
    Rhythm component-7 &Changes in shot lengths along the video& A & \cite{rhythmComponents} \\ \hline
    Brightness-7 &The value of \textbf{V} in the HSV space&V & \cite{audioAnalysis-8} \\ \hline
    Shot rate-7 & Rate of change in shot length&A & \cite{audioAnalysis-8,audioAnalysis-2} \\ \hline
    \end{tabular}
    \caption{\textbf{Audio-Visual Features} List of audio-visual multimedia features had been used for multimedia content analysis and a brief description along with the affective dimensions (V-Valence, A- Arousal) that have been linked to those features. Audio features are calculated using python library \emph{pyaudioanalysis} \cite{giannakopoulos2015pyaudioanalysis}. The length of the feature vector is mentioned next to the features name (in the first column).}
    \label{tab:MultiMediaFeatures}
\end{table*}

\subsection{Multimedia features based calculation of valence and arousal vectors}
\begin{figure*}[!h]
    \centering
    \includegraphics[width=\textwidth, height=0.8\textwidth]{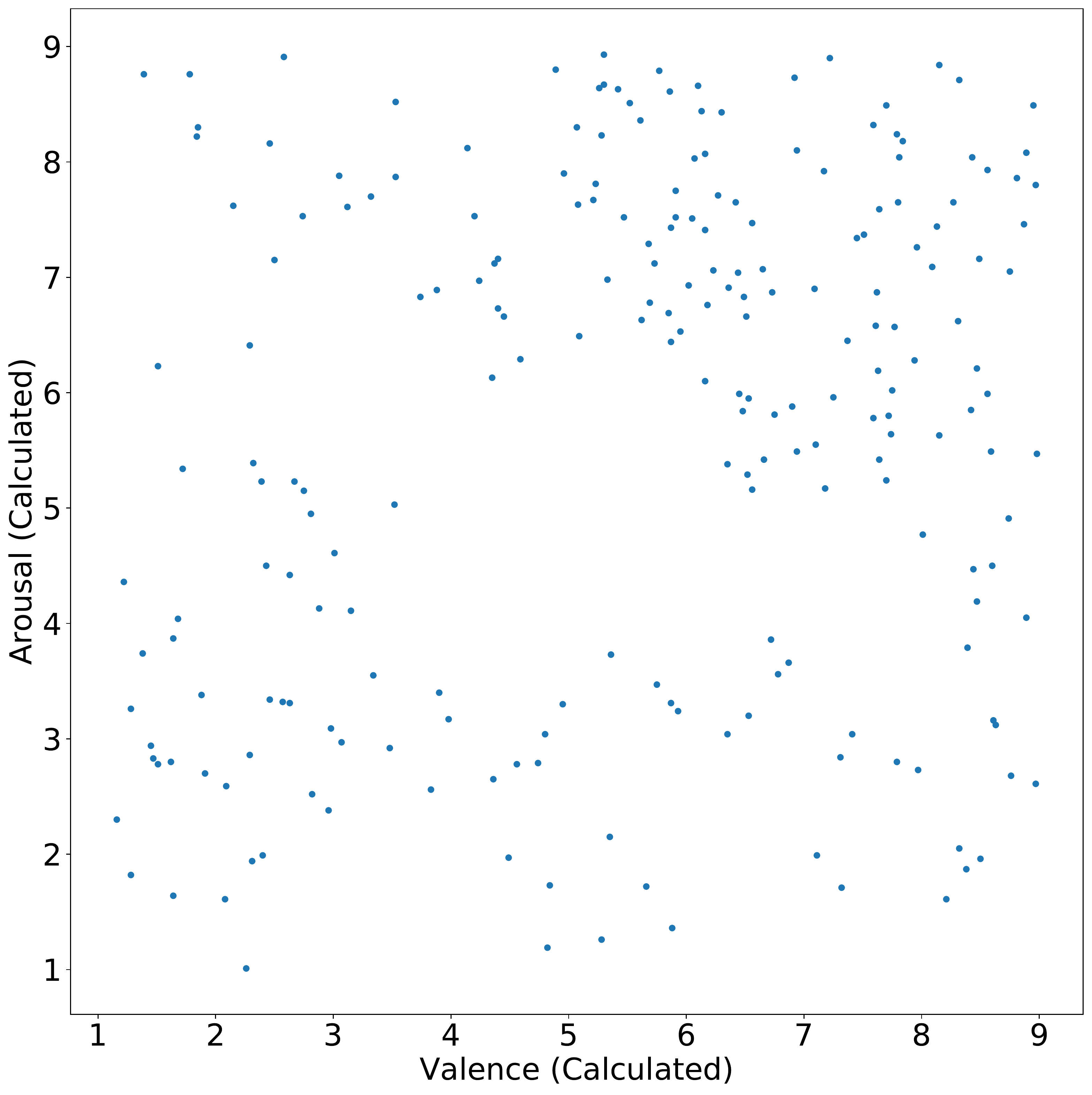}
    \caption{\textbf{Scatter plot for calculated valence and arousal values:} The scatter plot show the data points for 222 videos on the valence and arousal space. Valence (x-axis) and arousal (y-axis) were calculated from multimedia content analysis.}
    \label{fig:scatter222stimuli}
\end{figure*}

As described in the table-\ref{tab:MultiMediaFeatures} each audio-visual feature contributed a certain number of elements in the valence and arousal vector. The length of the valence and arousal vectors, considering only the audio features, were 38 and 64, respectively. On the other hand, the number of visual features used to calculate valence and arousal was the same (14 features). Dimensionality reduction on these valence and arousal vectors was performed using principal component analysis. Before applying PCA, the features were normalized to get rid of differences in the value range of different features. During the selection of the number of principal components, we made sure that the variance explained by the selected components was greater than 0.95. Hence, 15 and 5 principal components were considered for audio and video features, respectively. Each row of the transformed data represents a stimulus, and each column represents the selected principle dimension. The valence and arousal values from the transformed data were calculated (henceforth, this valence and arousal will be referred to as calculated valence and arousal to distinguish them from participants' ratings). The magnitude of the vector, comprised of elements along principal dimensions, is considered the calculated valence and arousal. The calculated valence and arousal values are re-scaled to the range 1 to 9 for comparison with subjective ratings. 
These calculated valence and arousal values were mapped on valence and arousal space as shown in the figure-\ref{fig:scatter222stimuli}. 

\subsection{Final set of emotional multimedia videos}
The seventy-two emotion words (see Appendix-1) were taken from the circumplex model presented in \cite{paltoglou2012seeing}. Two criteria were applied for the selection of emotion words. First, only emotional words which were away from the center of the circumplex model were selected (less than four and greater than six). The second criterion was related to the availability of valence and arousal ratings of emotion words in the list of English words provided by Warriner et al. \cite{warriner2013norms}. Each emotion word from the list of seventy-two emotion words had its corresponding mean and standard deviation of valence and arousal ratings studied by \cite{warriner2013norms}. Videos that had the calculated valence and arousal values (calculated from multimedia content analysis and re-scaled to range from one to nine) within the $mean\pm{sd}$ range of valence and arousal ratings (in a study by Warriner et al. \cite{warriner2013norms}), respectively, of any of the seventy-two emotion words, were selected. The mean and standard deviation of valence and arousal values by Warriner and colleagues \cite{warriner2013norms} gives the representative mean value and spread across the subjects. Hence, if the emotion word with the $mean\pm{sd}$ range of valence and arousal values by \cite{warriner2013norms} contained the calculated valence and arousal values, respectively, for any video, then it is assumed that the emotion word could be relatively more descriptive of emotional content than any other emotion word. Collectively, 222 stimuli were selected. These 222 selected stimuli were distributed in four quadrants of valence-arousal space as follows- HVHA:107, LVHA:35, LVLA:46, and HVLA:34. The scatter plot in figure-\ref{fig:scatter222stimuli} shows the distribution of 222 stimuli on valence-arousal space (in this plot, calculated valence and arousal values were used). Figure in Appendix-2 shows selected emotion words with means valence and arousal values (ratings provided in the study by \cite{warriner2013norms}.)

\subsection{Collection of subjective ratings}
\subsubsection*{Participants}
Students from the Indian Institute of Information Technology, Allahabad, India, were recruited through advertisements and presentations in some classes. Participants were credited for the laboratory course in exchange for their participation in the study. In addition, an appreciation e-certificate signed by PI was also provided. Some participants were excluded from the study due to their verbal confession of having history of mental disorders or if they were taking any medication for any potential conditions. They informed the experimenter about it (personally) and simply left the experiment room. An institutional review board, University of Allahabad (protocol code 2017-100 approved on Dec 8, 2017) approved the study. All the stimuli were reviewed by the review board and approved for their use in the study. Although English is not the native language of the participants, they have reported their preference for English music and movies. Also, for the participants, the instruction in schooling and college was English. Four hundred seven participants (female=64) with mean age 20.88 and standard deviation 2.27 participated in this study and rated 222 stimuli. 

\subsubsection*{Experiment}
The experimenter gave a presentation on the projector screen before the experiment to brief about it. The purpose of the presentation was not only to inform participants about the experiment paradigm but explaining them about scales and inform about the etiquette that needed to be maintained during the experiment. When all the doubts of participants were cleared, they were asked to sit in front of one of the computer in the alternate seats. The experiment environment was made more immersive by shielding the windows properly and switching off the lights. 

The experiment was performed in a big experiment hall with approx 45 participants participating in the experiment at the same time. As part of the training, participants were given three stimuli with the same paradigm they would follow during the experiment. Participants were helped in making them familiar with the paradigm and scaling. After the training session for each participant, they were allowed to participate in the experiment. 


\subsection{SVM classification of categories of rated valence and arousal from calculated valence and arousal vector}

We classified sets of two-hundred twenty-two stimuli into a different number of target classes. Input to our classifier was the calculated valence and arousal vectors in the previous step. In the 2-class classification, valence and arousal are classified as high or low valence and high or low arousal (mid-value five is considered threshold). On the other hand, with the 3-class division, we divided the valence and arousal scale into low, medium and high categories ($low<3.5,\;3.5<=mid<=6,\; high>6$). Since we had unbalanced data in target classes, we randomly selected a subset of samples from different categories. The length of each category-specific subset is equal to the length of the category with the minimum number of samples. Fifty iterations were run on different randomly sampled subsets. The mean and standard deviation of classification metrics are shown in the table-\ref{tab:CorrelationTest}. 

\section{Results}

\subsection{Representation of calculated valence \& arousal space}
Figure-\ref{fig:scatter222stimuli} shows the scatter plot for calculated valence and arousal values. These values were calculated as the magnitude of the vectors in the space of principal dimensions, as mentioned in the method section. These principle dimensions represented the dimension of maximum variations in the multimedia audio-visual features. The figure-\ref{fig:scatter222stimuli} clearly shows that the valence and arousal values are scattered around the whole space, validating our assumption that with the automatic method for initial video collection, a wide variety of affective contents can be covered.

\subsection{Participants ratings for 222 stimuli}
The complete table of participants' ratings is shown in Appendix-3. The table-\ref{tab:summaryIEEEAccess} shows ratings along all the dimensions are normally distributed. It means our dataset provides a balanced representation of positive and negative emotions. The data is also symmetrically distributed along the arousal axis. The summary of participants' ratings further verifies our approach of unbiased automatic collection and selection of affective stimuli. In addition, participants felt more arousal for stimuli from Indian cinema. For obvious reasons, participants liked the Indian stimuli more and reported more familiarity with the Indian stimuli. Like the overall ratings, for English stimuli, the mean and median for valence, arousal and dominance scale were around five depicting the symmetric distribution around the neutral rating of five. Whereas, for the Indian stimuli, the mean and median is slightly inclined towards higher values along valence and arousal dimensions. Liking and familiarity were also slightly skewed towards higher values for Indian stimuli. Overall, our method was able to extract a wide variety of positive and negative content without any experimenter biases.

\begin{table*}[h]
\centering
\resizebox{\textwidth}{!}{%
\begin{tabular}{c|ccc|ccc|ccc|ccc|ccc}
\textbf{} &
  \multicolumn{3}{c|}{\textbf{V\_mean}} &
  \multicolumn{3}{c|}{\textbf{A\_mean}} &
  \multicolumn{3}{c|}{\textbf{D\_mean}} &
  \multicolumn{3}{c|}{\textbf{L\_mean}} &
  \multicolumn{3}{c}{\textbf{F\_mean}} \\ \hline
 &
  Eng &
  Ind &
  OverA &
  Eng &
  Ind &
  OverA &
  Eng &
  Ind &
  OverA &
  Eng &
  Ind &
  OverA &
  Eng &
  Ind &
  OverA \\ \hline
\textbf{Min.}    & 1.99 & 1.75 & 1.75 & 3.11 & 3.43 & 3.11 & 3.38 & 2.95 & 2.95 & 1    & 1.52 & 1    & 1.54 & 2    & 1.54 \\
\textbf{1st Qu.} & 3.99 & 5.48 & 4.15 & 4.54 & 5.18 & 4.72 & 4.74 & 5.13 & 4.8  & 2.71 & 3.46 & 2.76 & 2.25 & 3.37 & 2.34 \\
\textbf{Median}  & 5.37 & 6.81 & 5.88 & 5.3  & 6.01 & 5.51 & 5.21 & 5.65 & 5.25 & 3.23 & 3.88 & 3.45 & 2.67 & 3.76 & 2.91 \\
\textbf{Mean}    & 5.24 & 6.08 & 5.47 & 5.29 & 5.99 & 5.48 & 5.18 & 5.6  & 5.29 & 3.19 & 3.66 & 3.32 & 2.76 & 3.66 & 3    \\
\textbf{3rd Qu.} & 6.46 & 7.36 & 6.81 & 6.08 & 6.63 & 6.25 & 5.66 & 6.2  & 5.76 & 3.68 & 4.23 & 3.92 & 3.2  & 4.12 & 3.57 \\
\textbf{Max.}    & 8.44 & 8.2  & 8.44 & 7.66 & 8.46 & 8.46 & 6.55 & 7.08 & 7.08 & 4.58 & 4.72 & 4.72 & 4.36 & 4.59 & 4.59 \\  \hline
\end{tabular}%
}
\caption{ Summary of ratings provided by participants along valence, arousal, dominance, liking and familiarity dimensions. There were 163 English videos and 59 videos from Indian cinema.
}
\label{tab:summaryIEEEAccess}
\end{table*}

\subsection{SVM classification of categories of rated valence and arousal}
We performed the classification of categories in valence and arousal using multimedia audio-visual features-based vectors of valence and arousal. Table-\ref{tab:CorrelationTest} shows that the calculated valence and arousal vectors using multimedia features have significant information, which can be used to predict the rated valence and arousal in terms of high vs low target categories. The mean and standard deviation of the classification metric is shown in the table-\ref{tab:CorrelationTest}. The results shown in table-\ref{tab:CorrelationTest} indicate that accuracy is better with the reduction in the number of film clips. The highest accuracy is observed for valence classification in the high and low classes. The lowest accuracy is observed when all stimuli are considered with 3-classes for arousal. The deviation of each iteration from the mean accuracy is very low, indicating that fluctuations in accuracy with different subsets of training and testing data are very low. 

\begin{table*}[h]
\centering
\caption{SVM classification of valence arousal ratings using calculated multimedia audio-visual features. 2-cat: high and low; 3-cat: high, medium and low.
}
\label{tab:CorrelationTest}
\resizebox{\textwidth}{!}{%
\begin{tabular}{lcccc}
\hline
& \multicolumn{2}{c}{Valence} & \multicolumn{2}{c}{Arousal} \\ \hline
& 2-Cat & 3-Cat & 2-Cat & 3-Cat \\ \hline
222 & $0.72\pm0.007$ & $0.68\pm0.01$ & $0.70\pm0.008$ & $0.67\pm0.03$ \\ \hline

\end{tabular}%
}
\end{table*}

\section{Discussion}
Our results show that the method we applied for the initial collection is able to reduce any experimenter biases. Moreover, we were able to get an equal representation of positive and negative stimuli in our data. The 222 stimuli contain English (163) and Indian (59) videos, making them suitable for use across Western and Indian cultures. However, our data has a limitation in that the videos were not annotated with the emotion labels. On the contrary, the good side is that if anybody wants to annotate some affective stimuli, the candidate stimuli collected using the objective and unbiased way will be at their disposal. Another advantage is that our data has no copyright issues. Hence, it can be easily downloaded by anyone interested in doing research with the affective multimedia stimuli. 

Many of the previous datasets used online mode for collecting the participants' ratings which suffers from problems including lack of focus, carelessness in following the experimenter's instructions, lack of control over the participant's environment, lack of control in audio and monitor settings and so on. All these factors contribute to very noisy data \cite{baveye2015liris}. On the contrary, our dataset is rated in a very controlled environment where participants were seated in a lab and were assisted for any problem. They were carefully watched during the experiment. The environmental condition were same for all the participants. They had similarities in their system's settings. Overall, ratings by participants in our dataset were less noisy than in the crowdsourcing-based collection.   

The dataset which we have reported in this article is the first dataset that participants from the Indian subcontinent validated. Moreover, the major advantage of our dataset is that it has stimuli from Western as well as Indian cinema, which makes it useful for research on Western as well as Indian population. The dataset recorded to date had the major limitation of their utilization in cross-cultural studies. In our work, we tried to address this problem in a novel way which has been rarely used in affective stimuli dataset collection. We believe that rather than picking the initial set of videos, if future research adopts our method, more cross-cultural datasets can be created.

\section{Conclusion}
In this work, we collected the first affective video stimuli dataset rated by Indian population. In addition, we tried to reduce the experimenter's selection biases by using YouTube API to crawl over the biggest user-created video platform for affective multimedia content. Consequently, we collected a wide variety of affective feelings, which may not be possible with the manual selection. Following our method, more cross-cultural datasets can be created so that affective computing and affective science can be understood in the context of cultural variation.

\bibliographystyle{unsrtnat}
\bibliography{references}  






\end{document}